\documentclass[aps,prl,twocolumn,longbibliography]{revtex4-1}
\usepackage{graphicx}
\usepackage{amsmath,amssymb}
\usepackage{mathrsfs}
\usepackage[normalem]{ulem}
\usepackage{color}
\usepackage{soul}
\usepackage{hyperref}
\usepackage[T1]{fontenc}
\usepackage[utf8x]{inputenc}
\usepackage{bm}
\usepackage[left]{lineno}

\newcommand{\angstrom}{\text{\normalfont\AA}}

\begin{document}


\title{Lattice dynamics localization in low-angle twisted bilayer graphene}

\author{Andreij C. Gadelha$^1$, Douglas A. A. Ohlberg$^1$, Cassiano Rabelo$^2$, Eliel G. S. Neto$^3$, Thiago L. Vasconcelos$^4$, João L. Campos$^1$, Jessica S. Lemos$^1$, Vinícius Ornelas$^1$, Daniel Miranda$^1$, Rafael Nadas$^1$, Fabiano C. Santana$^1$, Kenji Watanabe$^5$, Takashi Taniguchi$^5$, Benoit van Troeye$^6$, Michael Lamparski$^6$, Vincent Meunier$^6$, Viet-Hung Nguyen$^7$, Dawid Paszko$^7$, Jean-Christophe Charlier$^7$, Leonardo C. Campos$^1$, Luiz G. Cançado$^1$, Gilberto Medeiros-Ribeiro$^8$, Ado Jorio$^{1,2}$}

\affiliation{
$^1$Physics Department, $^2$Electrical Engineering Graduate Program, $^8$Computer Science Department, Universidade Federal de Minas Gerais, Belo Horizonte, MG 31270-901, Brazil.
$^3$Physics Institute, Universidade Federal da Bahia, Campus Universitário de Ondina, Salvador - BA, 40170-115 Brazil. 
$^4$Divis\~{a}o de Metrologia de Materiais, Inmetro, Duque de Caxias, RJ 25250-020 Brazil. 
$^5$National Institute for Materials Science (NIMS), 1-2-1 Sengen, Tsukuba-city, Ibaraki 305-0047, Japan.
$^6$Physics, Applied Physics, and Astronomy, Jonsson Rowland Science Center, Room 1C25 110 8th Street Troy, NY 12180, USA. 
$^7$Institute of Condensed Matter and Nanosciences (IMCN), University of Louvain (UCLouvain), Louvain-la-Neuve, Belgium. 
}

\begin{abstract}
\date{\today}
\textbf{A low twist angle between the two stacked crystal networks in bilayer graphene enables self-organized lattice reconstruction with the formation of a periodic domain. This superlattice modulates the vibrational and electronic structures, imposing new rules for electron-phonon coupling and the eventual observation of strong correlation and superconductivity. Direct optical images of the crystal superlattice in reconstructed twisted bilayer graphene are reported here, generated by the inelastic scattering of light in a nano-Raman spectroscope. The observation of the crystallographic structure with visible light is made possible due to lattice dynamics localization, the images resembling spectral variations caused by the presence of strain solitons and topological points. The results are rationalized by a nearly-free-phonon model and electronic calculations that highlight the relevance of solitons and topological points, particularly pronounced for structures with small twist angles. We anticipate our discovery to play a role in understanding Jahn-Teller effects and electronic Cooper pairing, among many other important phonon-related effects, and it may be useful for characterizing devices in the most prominent platform for the field of twistronics.} 

\end{abstract}

\maketitle

\begin{figure*}[!hbtp]
\centering
\centerline{\includegraphics[width=160mm]{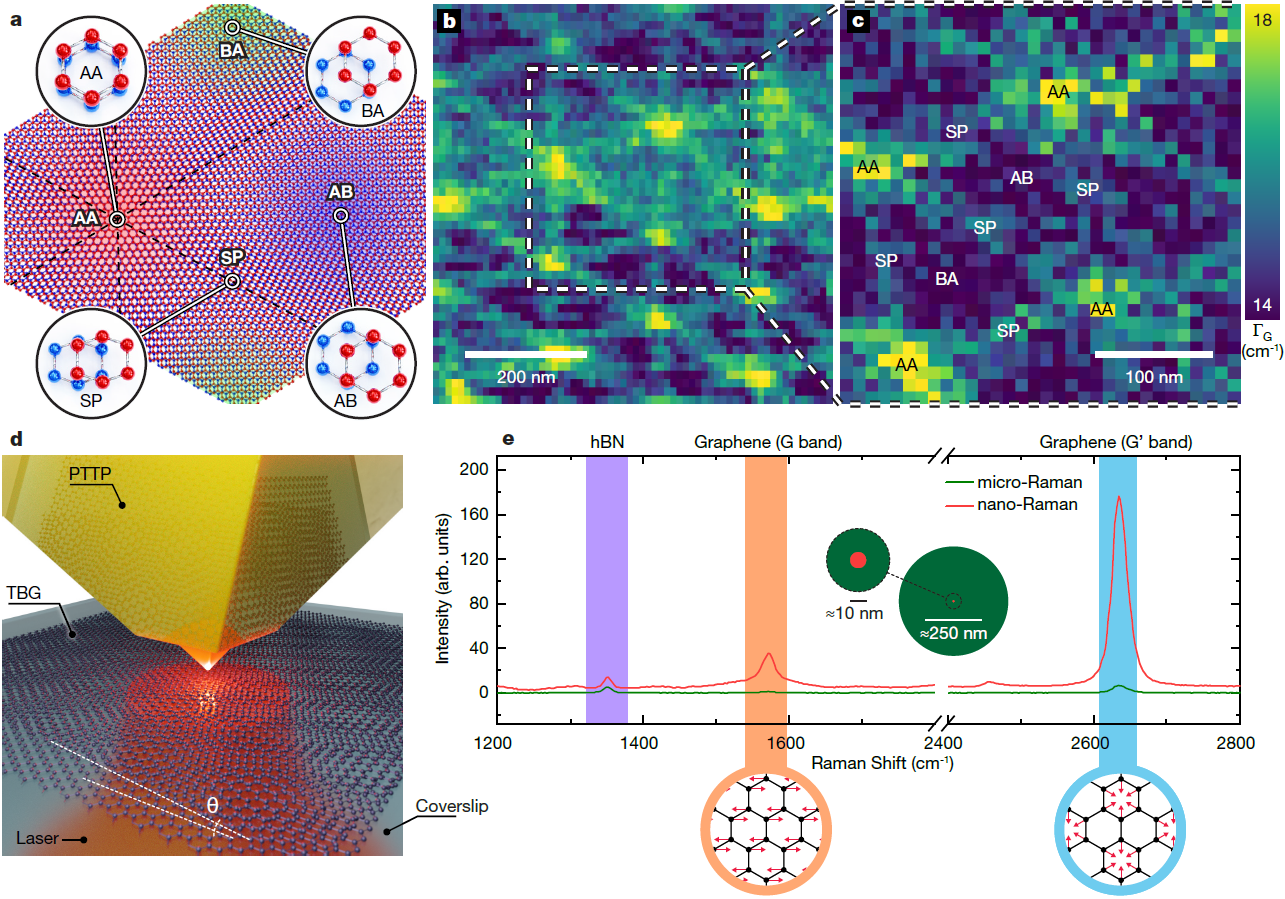}}
\caption{\label{fig:1}\textbf{Nano-Raman spectral imaging of a crystallographic superlattice in reconstructed twisted bilayer graphene}. {\bf a} Schematics showing neighbouring AB and BA-stacked domains, strain solitons (SP), and topological points (AA). {\bf b} Crystallographic hyperspectral image of a reconstructed twisted bilayer graphene based on the G$^{\prime}$ band nano-Raman intensity; {\bf c} Zoomed-in image from {\bf b} based on the G band nano-Raman linewidth ($\Gamma_G$). {\bf d} Schematics of the plasmon-tunable tip pyramid nano-antenna enhancing the Raman signal in a nanometric area, in the tip-enhanced Raman spectroscopy (TERS) configuration. {\bf e} Comparison between micro-Raman (green) and the nano-Raman (blue) spectra in the sample. Green and red circles schematically show the different illumination areas. The G and G$^{\prime}$ vibrational modes are depicted, and the peak from the hBN substrate.}
\end{figure*}

Graphite lattice dynamics have been widely studied for engineering the broadly utilized thermal and electrical properties of this semi-metal \cite{yoshimori1956theory}. Bilayer graphene, which in the so-called AB Bernal stacking represents the basic two-dimensional unit to build the three-dimensional graphite, has recently gained great attention because of its rich structural and electronic behavior when arranged with a small relative twist angle $\theta$ between the two layers. Below a threshold twist angle $\theta_c\sim 1^{\circ}$, the twisted bilayer graphene (TBG) undergoes an energetically favorable atomic reconstruction, entering the soliton regime for $\theta<\theta_c$ \cite{yoo2019atomic,oleg}. This equilibrium configuration possesses alternating AB and BA triangular stacking domains separated by shear solitons (SP stacking) in a hexagonal network \cite{alden2013strain,oleg}, with AA-stacked topological regions composing the vertices of the triangular areas  (see Fig.\,\ref{fig:1}{\bf a}). This reconstructed twisted bilayer graphene (rTBG) is a novel system, where local phenomena related to electronic/phononic reconstructions in addition to the morphology rearrangement \cite{oleg,lamparski2020soliton} are yet to be understood.

Experimentally, Raman spectroscopy, the inelastic scattering of light, has been a key technique to study the vibrational structure of graphite-related systems \cite{Tuinstra1970}, even gaining further importance for low-dimensional structures \cite{dresselhaus2010perspectives,Ferrari2013}, where inelastic neutron or X-ray scattering are difficult to use.
To be able to visualize the fine rTBG structure, however, a nano-Raman spectroscope, capable of resolving the optical information below the light diffraction limit, is necessary \cite{shao2019tip}. Figures\,\ref{fig:1}{\bf b},{\bf c} reveal a nano-Raman imaging of solitonic arrangements in a rTBG with 159\,nm superlattice period, which corresponds to $\theta = 0.09^{\circ}$. The specific nano-antenna of our nano-Raman setup, a plasmon-tunable tip pyramid \cite{Vasconcelos2018} shown schematically in Fig.\,\ref{fig:1}{\bf d}, is crucial to acquire the images illustrated in Figures\,\ref{fig:1}{\bf b},{\bf c}. It produces a local signal enhancement on the order of $3\times10^3$, generating a nano-Raman signal so intense that the micro-Raman response from the micron-sized illumination area becomes negligible (see Fig.\,\ref{fig:1}{\bf e}). The nano-Raman images are obtained at ambient conditions over regions of the bilayer that appear atomically flat and featureless in the surface topology images simultaneously obtained by the nano-antenna, which also functions as an atomic force microscope probe. Extremely clean rTBG samples without a top, capping hBN flake are also essential for high quality nano-Raman data. To produce such samples, we developed a new dry tear-and-stack method \cite{tear-and-stack}, based on a semi-pyramidal stamp that allowed the preparation of high-quality TBG flakes on a glass coverslip. These samples were extensively characterized by scanning probe microscopy techniques (SPM), including atomic force microscopy (AFM), scanning microwave impedance microscopy (sMIM) and scanning tunneling microscopy (STM). 

The particular arrangement in Fig.\,\ref{fig:1}{\bf b} was observed previously by transmission electron microscopy (TEM) \cite{alden2013strain,jiang2016soliton,yoo2019atomic} and nano-infrared spectroscopy \cite{photonictbg} techniques. The authors of these studies attribute the solitonic structure and soliton interceptions to shear strain solitons and topological AA points, respectively, based on the similarity between the observed superlattices and theoretical expectations for TBG reconstruction at low twist angles  (schematics in Fig.\,\ref{fig:1}{\bf a}). Here, the superlattice imaging is directly related to the local vibrational and electronic rTBG structure, since the nano-Raman spectroscopy probes the local atomic lattice vibration directly. The main Raman spectral signatures in graphene are due to the stretching of the C-C bonds (named G band, appearing at 1584~cm$^{-1}$) and the breathing motion of the hexagonal carbon rings (named G$^{\prime}$ band, symmetry-allowed overtone appearing at 2640~cm$^{-1}$), as assigned in Fig.\,\ref{fig:1}{\bf e}. Fig.\,\ref{fig:1}{\bf b} is a hyperspectral image based on the intensity of the rTBG Raman G$^{\prime}$ band, while Fig.\,\ref{fig:1}{\bf c} is a zoomed-in region of Fig.\,\ref{fig:1}{\bf b} based on the linewidth ($\Gamma_{\mathrm{G}}$) of the Raman G band, both without any statistical treatment or data filtering, just plotting the rough data. 


\begin{figure*}[!hbtp]
\centering
\centerline{\includegraphics[width=160mm]{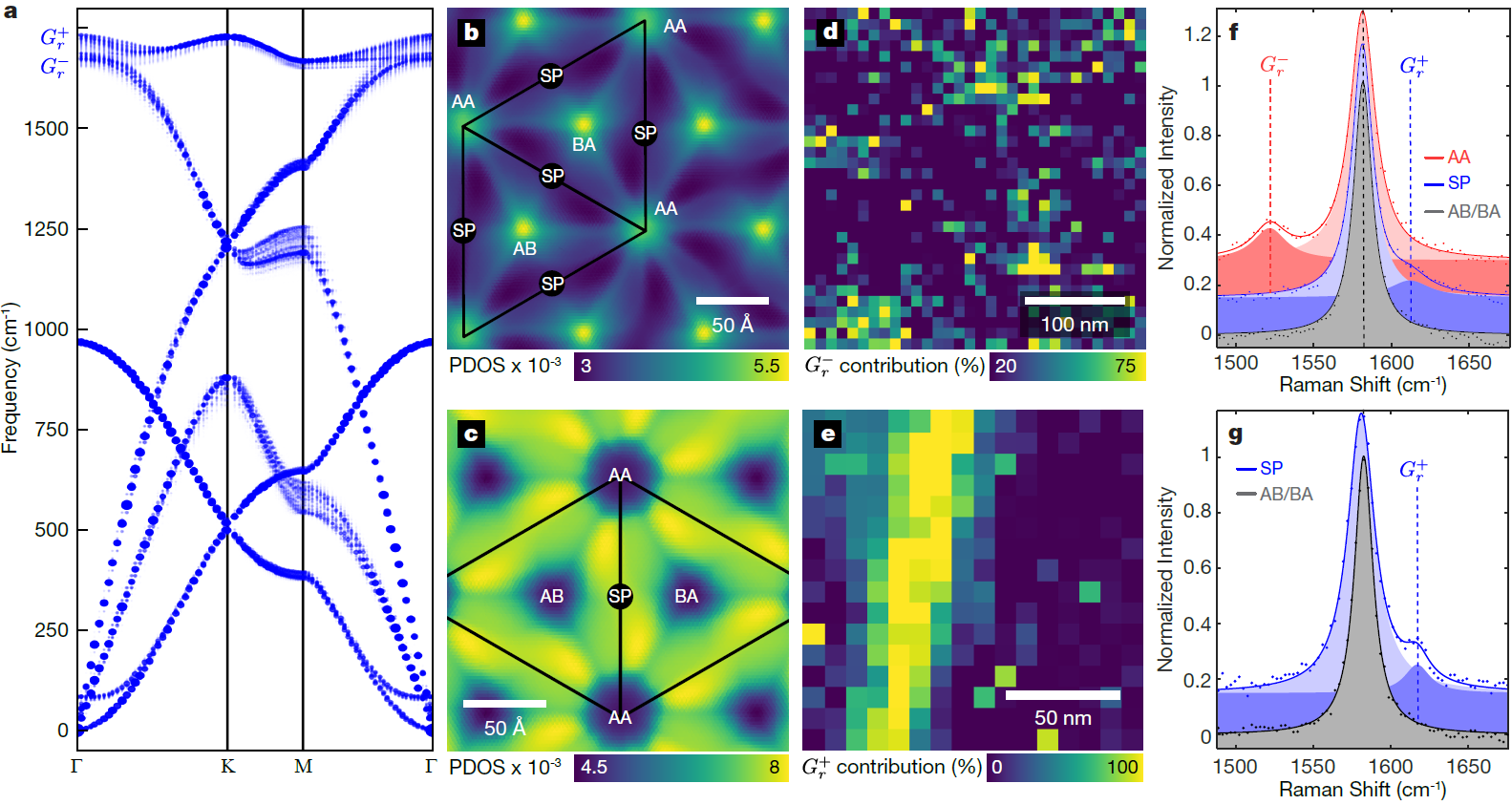}}
\caption{\label{fig:2}\textbf{Phonon structure and the nano-Raman spectral signature}. {\bf a} Theoretical phonon dispersion for Bernal AB-stacked twisted bilayer graphene (red) and for reconstructed twisted bilayers graphene (blue); {\bf b} and {\bf c} are the theoretically predicted spatial distributions of the phonon density of states for the lower ($G_r^-$) and higher ($G_r^+$) frequency optical phonons at the $\Gamma$ point, respectively. {\bf b} and {\bf c} are different in size and position with respect to each other, for better correlation to {\bf d} and {\bf e}.  {\bf d} and {\bf e} are the experimentally measured hyperspectral mapping of the $G_r^-$ and $G_r^+$ Raman peaks, shown in {\bf f} and {\bf g}, respectively. Data in {\bf d},{\bf f} comes from the same location as in Fig.\ref{fig:1}{\bf c}.
}
\end{figure*} 

The specific vibrational modes for the G and G$^{\prime}$ Raman bands (see inset to Fig.\,\ref{fig:1}{\bf e}) are not only different, but the scattering mechanisms that give origin to these Raman features \cite{dresselhaus2010perspectives,Ferrari2013} differ, as well. The G band is a first-order Raman active mode related to the double-denegerate high frequency optical phonon branch in graphene at the Brillouin zone center ($\Gamma$ point, see red lines in Fig.\,\ref{fig:2}{\bf a}). While the Bernal-stacked bilayer graphene exhibits a single phonon band (red G), the rTBG exhibits a splitting of the vibrations in several branches \cite{lamparski2020soliton}, two of which are predominant in the high frequency $\Gamma$ point, and the focus of our attention in the following. These phonon branches result from atomic reconstruction with the emergence of topological solitons. We show in Fig.\,\ref{fig:2}{\bf a} the (unfolded) phonon band-structure of rTBG (blue lines) with a twist angle of 0.987$^{\circ}$. 
In addition, we show in Fig.\,\ref{fig:2}{\bf b} and Fig.\,\ref{fig:2}{\bf c} the local phonon density of states around the lower phonon branch (G$_r^{-}$) and the higher phonon branch (G$_r^{+}$). These modes are, therefore, predicted to be localized in space, the lower frequency mode appearing more strongly at the AA regions, and the higher frequency mode preferentially in the soliton (SP) regions. A close inspection of the experimental Raman data shows the appearance of two satellite peaks next to the G band, also named here $G_r^+$ and $G_r^-$, appearing above and below the G band, and they are localized in space exactly as predicted by theory (see Fig.\,\ref{fig:2}{\bf b,d} and {\bf c,e}). Figs.\,\ref{fig:2}{\bf d},{\bf f} are data from the same location in Fig.\,\ref{fig:1}{\bf c}, and Fig.\,\ref{fig:2}{\bf d} renders the local intensity of the peak $G_r^-$ defined in Fig.\,\ref{fig:2}{\bf f}. Fig.\,\ref{fig:2}{\bf e} is a higher resolution imaging of a single soliton to better evidence the lower intensity $G_r^+$ peak, as shown in Fig.\,\ref{fig:2}{\bf g}. The frequency difference between the theoretically predicted $G_r^-$ and $G_r^+$ peaks is 45~cm$^{-1}$, while for the experimentally observed $G_r^-$ and $G_r^+$ peaks, it is {90}~cm$^{-1}$. This splitting is predicted to increase with decreasing twist angle \cite{lamparski2020soliton}, consistent with the different twist angles in experiment ($\theta=0.09^{\circ}$) and calculations ($\theta=0.987^{\circ}$), the later already representing our limit computational capability. 


\begin{figure*}[!hbtp]
\centering
\centerline{\includegraphics[width=160mm]{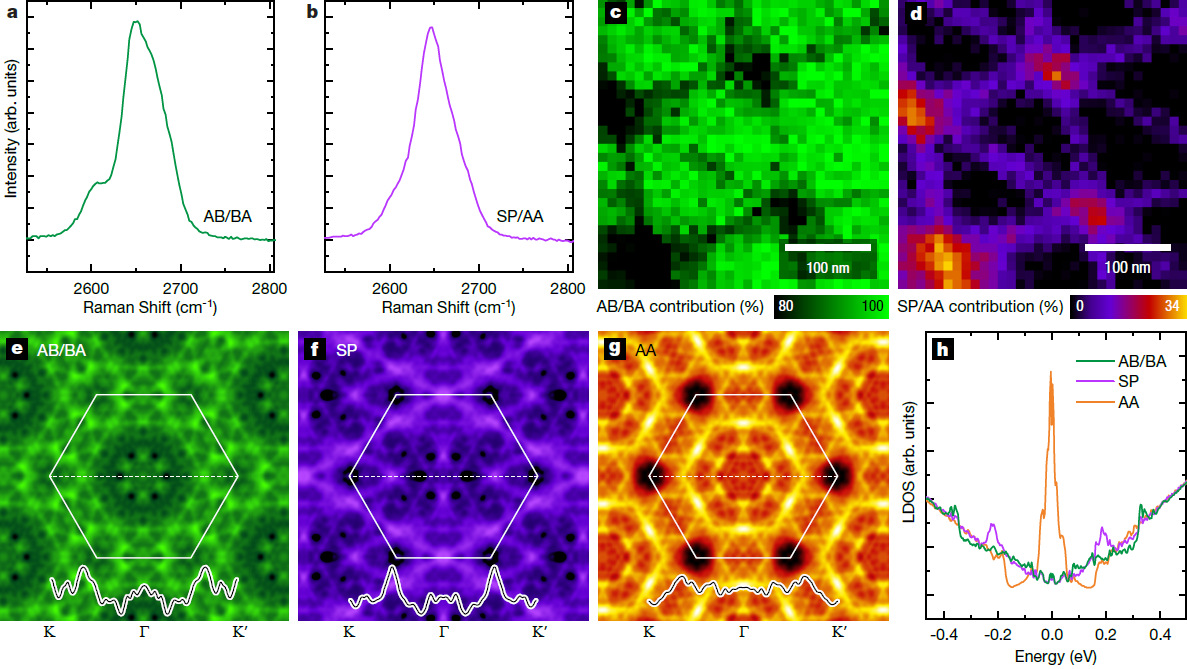}}
\caption{\label{fig:3}\textbf{The nano-Raman spectral signature (upper line) and the electronic structure (lower line)}. {\bf a} spectral Raman signature of the AB/BA-staked domains; {\bf b} corresponding data for the SP/AA domains; {\bf c} and {\bf d} are the spectral weight for the AB and SP/AA signatures, respectively, in the rTBG (same location as in Fig.\ref{fig:1}{\bf c}); {\bf e} to {\bf g} are the density of states at $E=-0.98$\,eV, plotted in momentum space, for the AB/BA, SP and AA regions, respectively. The insets show the line-trace for the DOS along the K-$\Gamma$-K$^{\prime}$ direction. {\bf h} DOS as a function of energy near the Fermi level for AB/BA, SP and AA regions.}
\end{figure*} 

The G$^{\prime}$ band, utilized to obtain Fig.\,\ref{fig:1}{\bf b}, is also related to the high-frequency optical phonon branch in graphene, but in the interior of the Brillouin zone, close to the K or K$^{\prime}$ points \cite{dresselhaus2010perspectives,Ferrari2013,thomsen2000double}. 
Figures\,\ref{fig:3}{\bf a} and {\bf b} plot exemplary nano-Raman spectra observed at an AB/BA region and at an SP region, respectively. We see no clear distinction between the soliton and AA region spectral signatures. The spectrum in Fig.\,\ref{fig:3}{\bf a} is typical of a Bernal-stacked bilayer, depicting four Lorentzian peaks \cite{ferrari2006raman}, thus confirming the AB-stacking structure, while the spectrum in Fig.\,\ref{fig:3}{\bf b} is different, displaying a unique shape for the SP/AA Raman signature. 
These spectral profiles were used to fit the G$^{\prime}$ hyperspectra shown in Fig.\,\ref{fig:1}, and the spectral weight of the AB/BA and SP/AA spectral information are plotted in Fig.\,\ref{fig:3}{\bf c} and Fig.\,\ref{fig:3}{\bf d}, respectively, evidencing clearly the rTBG structure. 

The G$^{\prime}$ lineshape is known to be sensitive to the electronic structure via electron-phonon coupling \cite{thomsen2000double,dresselhaus2010perspectives,Ferrari2013}. Moreover, the dependence on the number of Bernal-staking layers \cite{ferrari2006raman} and on twist-angle \cite{jorio2013raman} in bilayer has been established. Here, we show that in rTBG not only the phonon structure exhibits localization, but also the electronic structure, as predicted theoretically \cite{oleg}. Figures\,\ref{fig:3}{\bf e} to {\bf g} illustrate how the electronic density of states (DOS) in momentum space, at a fixed energy, changes locally for the rTBG. Calculations here are for a $\theta=0.505^{\circ}$ rTBG, our lowest achieved $\theta$ considering present DOS calculation technology. The fixed energy was chosen as $E_{DOS}=-0.98$\,eV, i.e. the energy for the valence electrons that are excited to the conduction band by our $E_{L}=1.96$\,eV excitation laser ($|E_{DOS}|= E_{L}/2$). These changes in the electronic structure are qualitatively reflected in the G$^{\prime}$ band spectral signature, see Figures\,\ref{fig:3}{\bf a} to {\bf d}. However, a clear quantitative analysis requires further theoretical development for addressing the electron-phonon coupling in these complex systems.


Changes in local DOS for rTBG also take place near the Fermi level, as shown in Fig.\,\ref{fig:3}{\bf h}, with further implications for the electron-phonon coupling and, consequently, for the nano-Raman imaging. G band phonons can be annihilated generating an electron-hole pair, and this mechanism decreases the overall G phonon lifetime, broadening the G peak in the Raman spectrum 
\cite{Lazzeri2006}.
A G band phonon quantum carries an energy of $\hslash \omega_{\rm{G}}\approx\rm{0.2\,eV}$. As a result, we expect that the presence of DOS peaks at $\pm 0.2$\,eV from the Fermi level in Fig.\,\ref{fig:3}{\bf h}, which are found to be preserved even for smaller twisted angles, causes a broadening of the rTBG G band in the SP (green) and AA (blue) regions. These predictions agree with the results shown in Fig.\ref{fig:1}{\bf c}, notably in the AA regions, where the G band is wider due to the DOS peak also at the Fermi level (see Fig.\,\ref{fig:3}{\bf h}). 


\begin{figure}[!hbtp]
\centering
\centerline{\includegraphics[width=89mm]{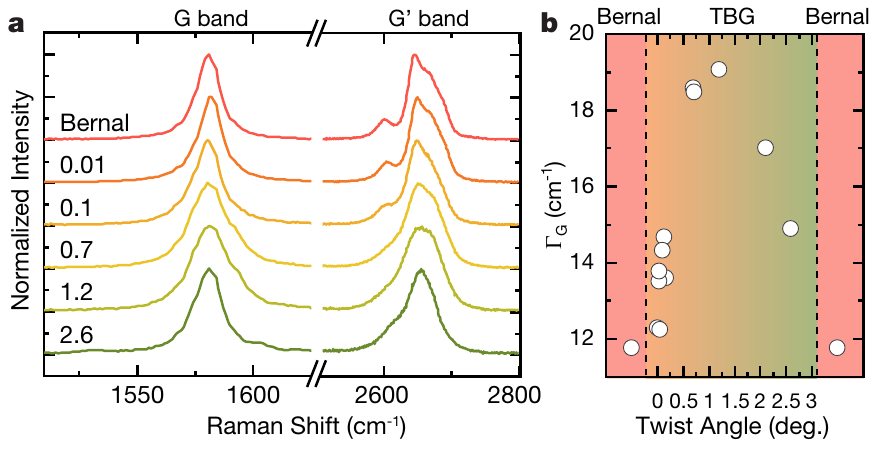}}
\caption{\label{fig:4}\textbf{Micro-Raman spectral fingerprint for different twisted bilayer graphene at different twist angles.} {\bf a} brings the G and G$^{\prime}$ band spectra and {\bf b} the full-width at half maxima of the G peak ($\Gamma_{\rm{G}}$). The rTBG $\theta$ values were measured by high-resolution sMIM. The spectrum for Bernal AB-stacking was measured for reference.}
\end{figure} 

Finally, to link the nano-Raman scattering experiments presented here with the usual micro-Raman spectroscopy characterization of twisted bilayer graphene, Fig.\,\ref{fig:4}{\bf a} shows the micro-Raman G and G$^{\prime}$ spectra for different twist angles between 0.01 and 2.6 degrees. We also plot in Fig\,\ref{fig:4}{\bf a} the spectra for the Bernal-bilayer graphene, for reference, and Fig.\,\ref{fig:4}{\bf b} shows the full-width at half maximum (FWHM) for the observed G band ($\Gamma_{\rm{G}}$). An increase of $\Gamma_{\rm{G}}$ when decreasing the twist angle below $\theta = 5^{\circ}$ was reported \cite{RIBEIRO2015138}, but here we see evidence that it reaches a maximum near the magic angle ($\theta \sim 1.1^{\circ}$) \cite{cao2018unconventional}, and it decreases for lower twist angle values, back to the reference Bernal AB-stacking value at $\Gamma_{\rm{G}}=12$\,cm$^{-1}$, see Fig.\,\ref{fig:4}{\bf b}. Considering reconstruction regime for {$\theta < 1.2^{\circ}$ \cite{oleg}}, when the angle decreases, the ratio between the AA/SP and AB/BA areas decrease, and the TBG G band tends to the corresponding AB version, see Fig.\,\ref{fig:4}{\bf b}. It is interesting, however, to find $\Gamma_{\rm{G}}$ as high as 18\,cm$^{-1}$ near the magic angle. This value is higher than graphene at the charge neutrality point, where the electron-phonon coupling is maximum \cite{Lazzeri2006}.
$\Gamma_{\rm{G}}$ is directly related to the electron-phonon coupling, 
and this result indicates a peak in the electron-phonon coupling near the magic angle, an evidence of the possible role of phonons in graphene superconductivity. Future low-temperature gate-doping experiments \cite{Lazzeri2006} could clarify the importance of electron-phonon in $\Gamma_{\rm{G}}$, as contrasted to other possible structural effects \cite{cocemasov2013phonons,lamparski2020soliton}. In that way, it is possible to use the micro-Raman spectra to evaluate twist-angle disorder \cite{uri2020mapping} in rTBGs and for searching regions close to the magic angle.


In closing, it is important to stress that twisted bilayer graphene (TBG) has drawn increasing attention since the discovery of strongly correlated phenomena, such as unconventional superconductivity 
\cite{cao2018unconventional}.
Here we bring valuable local lattice vibration and electron-phonon coupling information into this problem, which should play a role in understanding Jahn-Teller effects \cite{angeli2019valley} and electronic Cooper pairing \cite{wu2018theory,wu2019phonon,lian2019twisted}. This achievement was only conceivable because of the new tear-and-stack TBG preparation method and the use of plasmon-tunable tip pyramids \cite{Vasconcelos2018} for tip-enhanced Raman spectroscopy. In general, this work provides an important tool for the implementation of twistronics  
with an extra degree of manipulation \cite{mele2010commensuration} of many quantum properties and exotic phenomena absent in pristine graphene, like ferromagnetism 
\cite{Sharpe605}, anomalous quantum Hall effect \cite{Serlin900}  and large linear-in-temperature resistivity \cite{Polshyn2019}, and it may be useful for characterizing devices \cite{uri2020mapping}. All these new aspects come with the ability, presented here, to observe a crystallographic Moiré pattern using visible light.

\bibliographystyle{unsrt}

\newpage

\section{Supplementary Material}

\subsection{Experimental Details}

\begin{figure*}[hbtp]
\centering
\includegraphics[width=160mm]{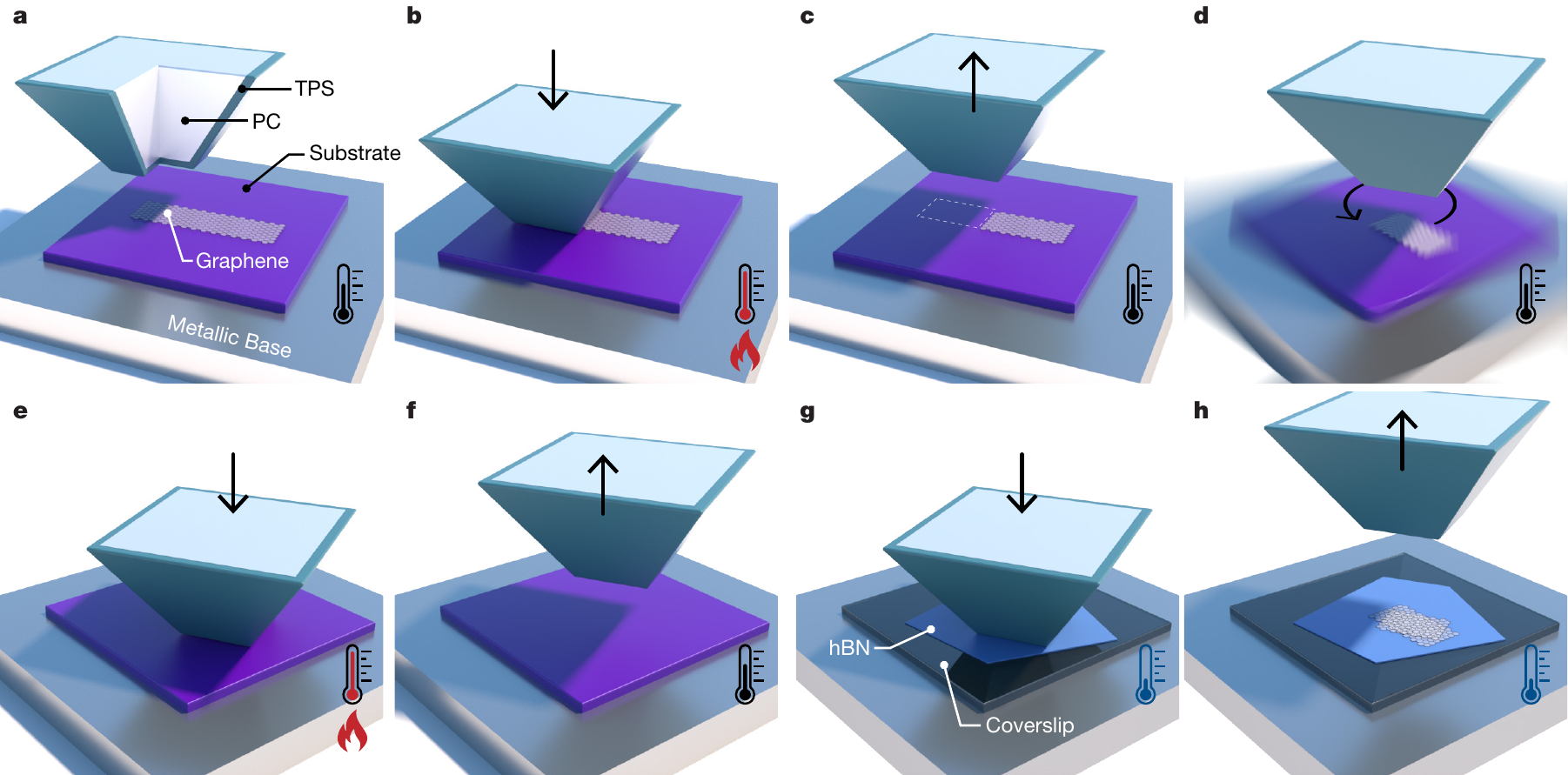}
\caption{\textbf{Schematics for th sample preparation procedure}. We cover the newly developed Polydimethylsiloxane (PDMS) tear-and-stack pyramid stamp (TPS) with a polycarbonate (PC) sheet and align the TPS edge with the middle of a graphene flake, see \textbf{a}. Next, we make contact between the TPS and graphene, followed by a temperature ramp from 70~$^{\circ}$C to 80~$^{\circ}$C, see \textbf{b}. We then wait for the system to cool down, reaching the temperature 70~$^{\circ}$C similarly to the pick-up method \cite{pick}, removing a piece of the graphene flake, see \textbf{c}. Following, we rotate the base precisely, see \textbf{d}, and stack the two parts of graphene together, forming the twisted bilayer graphene (TBG), see \textbf{e}. We do the previous temperature ramp and cool-down procedure again to remove the remaining graphene piece, see \textbf{f}. Next, we make contact between the TBG and a flat and clean boron nitride (hBN) flake at room temperature, see \textbf{g}. The Van-der-Waals interactions between them are strong enough for the hBN to pull out the TBG from the TPS, see \textbf{h}.}
\end{figure*}

\begin{figure*}[hbtp]
\centering
\includegraphics[width=160mm]{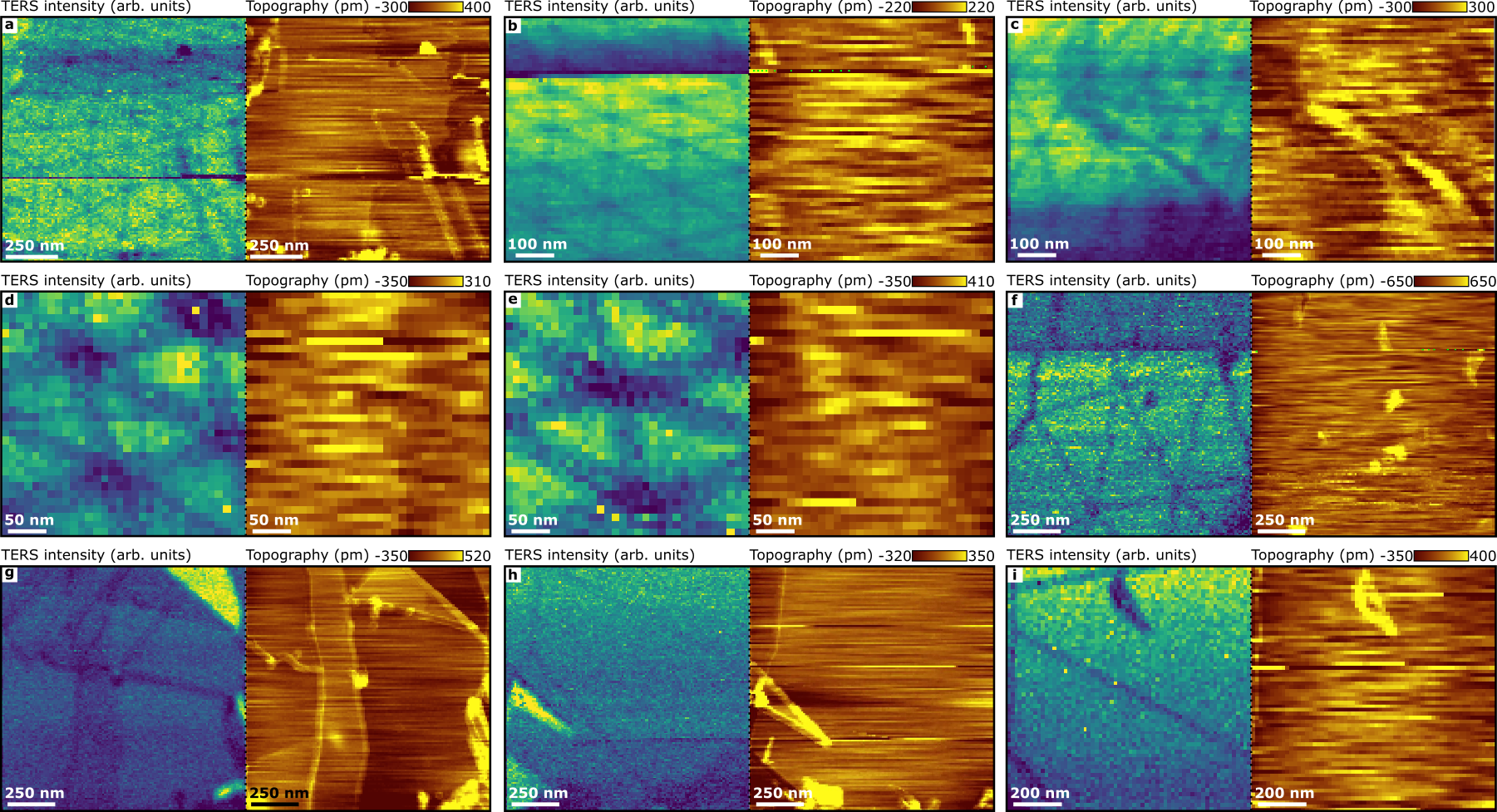}
\caption{\textbf{TERS imaging of nine different Moiré structures from reconstructed twisted bilayer graphene (rTBG)}. \textbf{a-i} display the TERS images from a diversity of Moiré structural formations based on the G$^{\prime}$ band intensity, on the left and, on the right, the simultaneously obtained atomic force microscopy (AFM) image. Notice the solitonic structures are only observable in the TERS image (darker blue lines), and they are absent in the AFM images. The contrast discontinuities in figures \textbf{b} and \textbf{f} are due to the realignment of the tip with the laser during the scan.}
\end{figure*}

\begin{figure}[hbtp]
\centering
\includegraphics[width=89mm]{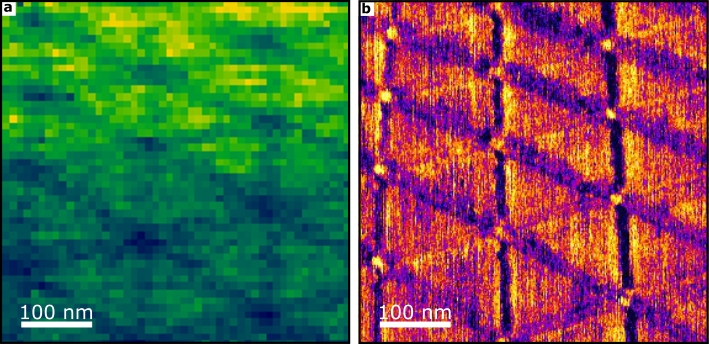}
\caption{\textbf{Multi-technique structural characterization of rTBG}. \textbf{a} and \textbf{b} show, respectively, tip-enhanced Raman scattering (TERS) and Scanning Microwave Impedance Microscopy (sMIM) images of the same rTBG region.}
\end{figure}

\newpage
\subsection{Electronic Calculations}

To compute the electronic structure of reconstructed twisted bilayer graphene, the relaxed structure obtained by the phonon-model calculation was taken into account and the p$_{\mathrm{z}}$ tight-binding (TB) Hamiltonian

\begin{equation}\label{eq1}
H_{tb}=\sum_{n}\mu_{n}a_{n}^{\dagger}a_{n}+\sum_{n,m}t_{nm}a_{n}^{\dagger}a_{m}+h.c
\end{equation} was then employed. In particular, the hopping terms $t_{nm}$ between neighboring atoms are given by the standard Slater-Koster formula

\begin{equation}\label{eq2}
t_{nm}=\mathrm{cos^{2}}\phi_{nm}V_{pp\sigma}\left(r_{nm}\right)+\left( 1-\mathrm{cos^{2}}\phi_{nm}\right)V_{pp\pi}\left(r_{nm}\right)
\end{equation}
where the direction cosine of $\overrightarrow{r}_{nm}=\overrightarrow{r}_{m}-\overrightarrow{r}_{n}$ along Oz axis is $\mathrm{cos}\phi_{nm}=z_{nm}/r_{nm}$. Similarly as in Ref. \cite{Tramblyde2012}, the distance-dependent Slater-Koster parameters are determined as
\begin{equation}\label{eq3}
V_{pp\pi}\left(r_{nm}\right)=V_{pp\pi}^{0}\mathrm{exp}\left[q_{\pi}\left(1-\frac{r_{nm}}{a_{0}}\right)\right]F_{c}\left(r_{nm}\right)
\end{equation}
\begin{equation}\label{eq4}
V_{pp\sigma}\left(r_{nm}\right)=V_{pp\sigma}^{0}\mathrm{exp}\left[q_{\sigma}\left(1-\frac{r_{nm}}{d_{0}}\right)\right]F_{c}\left(r_{nm}\right)
\end{equation}
with a smooth cutoff function $F_{c}\left(r_{nm}\right)=\frac{1}{1+\mathrm{exp}\left(\frac{r_{nm}-r_{C}}{\lambda_{C}}\right)}$.

\begin{figure*}[!hbtp]
	\centering
	\includegraphics[width=160mm]{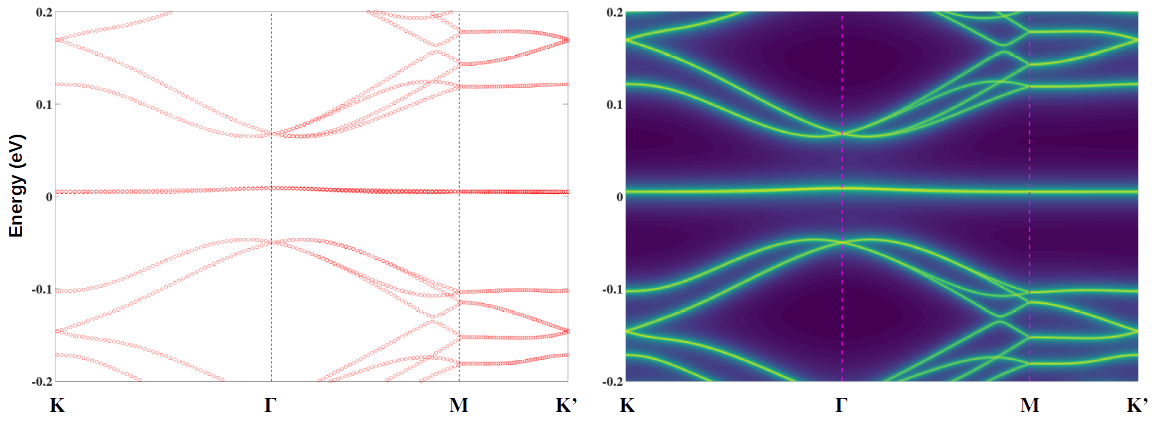}
	\caption{\label{figele}\textbf{Electronic bandstructure at the first magic angle 1.121$^{\circ}$.}
			Modeled by direct diagonalization (left) and Green´s function techniques (right).}
\end{figure*}

To model more accurately the first magic angle (with a zero-energy flat band) experimentally observed at $\approx$~1.1$^{\circ}$ \cite{cao2018unconventional} (see Fig.~\ref{figele}), our calculated TB parameters were obtained by slightly adjusting those reported in \cite{Tramblyde2012}. In particular,
\begin{eqnarray*}
V_{pp\pi}^{0}&=&-2.7\,\mathrm{eV} \\
V_{pp\sigma}^{0}&=&375\,\mathrm{meV} \\
\frac{q_{\pi}}{a_{0}}&=&\frac{q_{\sigma}}{d_{0}}=2.218\,\angstrom^{-1} \\
a_{0}&=&1.42\,\angstrom,\,d_{0}=3.43\,\angstrom \\
r_{c}&=&6.14\,\angstrom,\\
\lambda_{c}&=&0.265\,\angstrom
\end{eqnarray*}
The onsite energy is fixed at O$_{n}$ = --783.75~meV.

Basically, the electronic bandstructure and corresponding electronic quantities can be computed by diagonalizing the Hamiltonian (equation \ref{eq1}). However, diagonalization calculations in the case of small twist angles ($\ll$ the first magic one) are very numerically challenging due to a huge number of atoms in their supercell. To solve this numerical issue, we employed an alternative method based on the Green´s function techniques computing the real-space and momentum dependent local density of electronic states LDOS($E$,\textbf{r},\textbf{k}). Actually, this function represents its high peaks only when the electron energy satisfies $E \equiv \omega_{n}
(\mathbf{k})$ ($\omega_{n}
(\mathbf{k})$ is the n-th eigenvalue of the Hamiltonian (equation \ref{eq1}) at the momentum \textbf{k}). This thus allows to model the electronic structure using the real-space/momentum maps of LDOS($E$,\textbf{r},\textbf{k}) (see the example in Fig.~\ref{figele}). The main advantages of this Green´s function method, compared to direct diagonalization, include: (i) minimizing the size of calculated matrices by using recursive techniques \cite{THORGILSSON2014256} and (ii) efficient for calculating both global and local electronic properties. The former enables calculations of extremely small twist angles, i.e., large Moiré superlattices.


\end{document}